\documentstyle[12pt,epsf]{article}
\def\epm{e$^+$e$^-$}
\def\emm{e$^-$e$^-$}
\def\emg{e$^-\gamma$~}
\def\gg{$\gamma \gamma$~}
\def\E{{\cal E}}
\begin{document}
\begin{flushright}
\small{
SLAC-PUB-7426 \\
HUPD-9707 \\
April, 1997 \\
}
\end{flushright}
\begin{center}
{\LARGE Simulations of the Interaction Region \\
in a Photon-Photon Collider \\ }
\vspace{5 mm}

 {\large P.~Chen$^{\rm a}$, T.~Ohgaki$^{\rm b}$, A.~Spitkovsky$^{\rm c}$, 
  T.~Takahashi$^{\rm b}
 \footnote{corresponding author. e-mail address: tohrut@kekux1.kek.jp}
 $ and K.~Yokoya$^{\rm d}$ \\ }
 \vspace{3 mm}
  $^{\rm a}$Stanford Linear Accelerator, Stanford University,
          Stanford, CA94309, USA \\
  $^{\rm b}$Department of Physics, Hiroshima university, 
	    Higashi-Hiroshima, 739, Japan \\
  $^{\rm c}$Department of Physics, University of California at Berkeley,
	  Berkeley, CA 94720, USA \\
  $^{\rm d}$KEK-National Laboratory for High Energy Physics, 
	 1-1 Oho, Tsukuba, 305, Japan \\
\end{center}
%

\begin{abstract}
The status and initial performance of a simulation program CAIN
for interaction region of linear colliders is described.
The program is developed to be applicable for  \epm, \emm, \emg
and \gg linear colliders.
As an example of an application, simulation of a \gg collider
option of NLC is reported.
\end{abstract}
\section{Introduction}

 As additional options of \epm linear colliders, 
 feasibility studies for \emm , \emg  and \gg
 colliders have become active in recent years.
 For linear colliders, detailed knowledge of beam-beam
 interaction is important
 to estimate backgrounds in the detector as well as
 to calculate realistic luminosity. 
 In order to study these effects, 
 a Monte Carlo simulation program for beam-beam interaction
 in \epm colliders, ABEL (Analysis of Beam-beam Effects in Linear colliders),
 was developed \cite{ABEL}.
 The original ABEL included 
 beam disruption and beamstrahlung \cite{pairs} and 
 was later modified (ABELMOD) \cite{ABELMOD} to include electron-position 
 pair creation which is potentially  a serious source of background
 to the detector \cite{CAIN0}.
 
 For \emm, \emg 
 and \gg colliders, a similar kind of simulation is necessary 
 to understand beam-beam interaction, but the situation 
 is much more complicated due to the necessarily complex schemes of
 the beam collision.
 In \emg and \gg colliders, an intense laser beam is 
 flashed on the incoming electron beam 
 just before the interaction point to convert high energy electron 
 beam to high energy photon beam by Compton scattering \cite{Ohgaki}.
 A simulation program for these colliders is required to implement 
 laser-electron interaction at the conversion point which is 
 typically on the  order of cm upstream from the interaction point.
 In addition to the conversion point, the simulation of beam-beam
 interaction at the interaction point is more complicated than in
 the \epm collider case, since 1) the initial state includes
 not only electrons and positron but also photons, 2) as a consequence of 
 Compton interaction at the conversion point, 
 electron and photon beams have wider spectrum in both energy and spatial
 distribution than those of \epm colliders.
 According to these features,
 it is necessary to develop a new simulation program for 
 beam-beam interaction.
 To meet this requirement, a project to write a new
 comprehensive beam-beam simulation  program named CAIN (Conglomerat d'ABEL
  et d'Interactions Non-lineares) \cite{CAIN0} was launched
  with the intention to include Compton and Breit-Wheeler
 processes at conversion point, transport of particles from 
 conversion point to interaction point and interaction of all
 possible combinations of electrons/positrons and photons at the
 interaction point. 
 By  using the Compton scattering part of  CAIN,
 the effect of Breit-Wheeler process in 
 a photon-photon collider is discussed  
 in previous paper\cite{Ohgaki}.

 In this paper, we report the first version of the 
 comprehensive simulation program which can treat
 conversion, transportation and interaction region in  
 a single framework
 and  is applicable
 for handling of all 4 types of \epm , \emm , \emg  and  \gg colliders.
 As examples of the simulation,
 differential luminosity distribution of
 \gg, \emm  and \epm option of NLC
 is described.

\section{Structure of CAIN (version 1.0)}

 A schematic of a \gg collider is 
 illustrated in fig.\ref{fig:schematic}.
 An intense laser pulse is flashed on the 
 electron beam at the conversion point (CP) where
 high energy photon beams are produced by 
 Compton scattering. 
 Photons and electrons coming out from 
 CP are transported for O(cm) to 
 the interaction point (IP).
 At the transport region (TP), spent electrons 
 from CP may be swept out by 
 an external magnetic field or possibly plasma
 lens to avoid the electron collision at the IP
 Electrons and photons that are transported to the IP 
 collide with  positrons  and photons from another beam.
 
 Corresponding to the scheme of the \gg collider,  
 simulations in CAIN are divided in three modules
 CP, TP and IP as
 illustrated in fig.\ref{fig:structure}.
 In this figure, particles and processes included in each step
 are shown as well. 

 At the CP, Compton and Breit-Wheeler
 interactions between laser and  electron beams are simulated.
 First of all, an electron bunch is divided into 
 given number of macro particles
 and initial position and momentum of each macro particle are
 calculated from beam twiss parameters. 
 In a typical linear collider, the
 number of electrons per bunch is 
 O($10^{10}$). 

 A typical simulation uses 10,000 macro particles, with each one 
representing O($10^6$) electrons.
 The transverse and longitudinal coordinates in the simulation space
 are subdivided into cells and macro particles are assigned to these cells 
according to their position. 

 The time in the simulation is also divided into steps.
 In each time step, the probability of Compton scattering is 
 calculated for each macro particle according to the laser intensity,
 and a Compton scattered photon is generated according to the probability.
 The local laser intensity at each cell is calculated from given 
 laser parameters by taking into account diffraction effect, i.e:,
 $$
 \sigma _r^L(z)=\sigma _0^L\sqrt {1+\left( {{z \over {Z_R}}} \right)^2}
 $$
 where $\sigma^L_0$ and $\sigma^L_r(z)$ are RMS spot sizes 
 at the focal point and at distance $z$ from the focal point of
 the laser. 
 $Z_R$ is the Rayleigh length of the laser which is  defiend as;
 $$
 Z_R={{4\pi (\sigma _0^L)^2} \over {\lambda _L}}
 $$
 where $\lambda_L$ is wave length of the laser.
 If the Compton scattering event is generated, the new 
 photon is created and  momentum of the scattered electron 
 is modified. Such an electron can still interact with
 the laser in the following time steps.
   
 As is described later, the primary consideration in selecting 
 the laser parameters is keeping the  effect of nonlinear QED processes
 to a minimum. However, it is impossible to avoid such 
 effects completely when high luminosity is required.
 The nonlinear Compton processes can be  expressed as 
$$
e+n\gamma (laser)\to e+\gamma 
$$
where $n\gamma (laser)$ indicates that more than one laser
photons are absorbed coherently by a single electron.
Since this process accounts for higher tails in scattered photon spectrum 
and lowers peak energy of photon spectrum due to the increase of 
effective electron mass \cite{volkov},
it has to be kept small to get good photon beams. 
For the purpose of \gg and  \emg \ collider application, 
proper treatment of helicity of electron and laser beam is 
essential since produced photon energy spectrum depends
on the helicity state of incoming electrons and photons. 
In the simulation, cross section formula by Tsai \cite{tsai}
is used in which the polarization of electrons and laser is
taken into account in nonlinear QED calculation.  
The nonlinear Breit-Wheeler process can be written as
$$
\gamma(Compton) +n\gamma (laser)\to e^+e^-
$$
where more than one laser photons are absorbed by 
(Compton scattered) high energy photon and 
generate electron positron pairs. 
This process produces electron-positron pairs 
even if the center of mass energy of $\gamma(Compton) \gamma(laser)$
system is lower than \epm \ threshold
and could be an additional source of backgrounds
in high laser field environment.
We also adopt the formula by Tsai \cite{tsai} for this process.

 The transport process takes care of drift of
 spent electrons, Compton photons and electron 
 positron pairs that come out from CP.
 Photons are simply drifted to IP
 according to their angular divergence.
 For electrons and positrons, however,
 it is possible to insert a sweeping  magnet
 to deflect them from IP.
 In the version 1.0 of CAIN, a sweeping magnet 
 and synchrotron radiation in the sweeping process
 is included by a classical radiation treatment.
 Since the strength of the sweeping magnetic
 field is on the order of 1 Tesla, the critical energy of 
 synchrotron radiation is low enough to be 
 treated as the classical radiation. 
 Synchrotron radiation photons are added to the total 
 photon population to be inputed to the interaction region.

 IP phenomena are simulated in the same way as done in ABELMOD 
 \cite{ABEL,ABELMOD}. 
 In fact,
 a reworked version of ABELMOD serves as an interaction region module
 in CAIN 1.0 to simulate disruption of electron beams, generation of
 beamstrahlung, and production of low energy pairs. 
 The difference from ABELMOD is that CAIN needs to  take care of
 mixture of  electrons/positrons and photons as its initial state, while 
 only electron and positron beams could be used in ABELMOD. 
 Thus ABELMOD was modified to treat externally
 supplied photons and internally generated beamstrahlung
 photons on equal footing.
 In every time step in the interaction region 
 CAIN collects total and differential luminosities of 
 \epm \ or \emm \ as well as \gg and e$\gamma$ 
 luminosities that are available  for graphical display after the simulation. 
 
\section{Case Study: \gg, \emm Collisions in NLC}
\subsection{\gg Collisions}

Simulations of \gg collisions were performed with 
the  reference  parameters
for a \gg collider option of  NLC \cite{NLC} summarized in 
Table 1.

\begin{table}
\begin{center}
\caption{Parameters for a photon-photon collider} 
\begin{tabular}{|ll|}
\hline
{Electron beam parameters} & \\
\hline
  Beam energy                      & $\E_b$=250{\rm GeV} \\               
  Number of Particles per bunch    & $N=0.65 \times$ $10^{10}$  \\          
  Repetition rate                  & $f_{rep}=180{\rm Hz}$     \\
  Number of bunches per pulse      & $n_b=90$  \\
  Bunch length                     & $\sigma_z$=100$\mu$m \\     
  Bunch sizes (C.P.)               & $\sigma_x$=718nm \\
                                   & $\sigma_y$=91nm \\
  Bunch sizes (I.P.)               & $\sigma_x$=71nm \\
                                   & $\sigma_y$=9.1nm \\
  Beta functions (I.P.)            & $\beta_x$=0.5mm \\
                                   & $\beta_y$=0.5mm \\               
  Emittances & $\gamma\varepsilon_x$=5.0 $\times$ $10^{-6}$ m$\cdot$rad \\
             & $\gamma\varepsilon_y$=8.0 $\times$ $10^{-8}$ m$\cdot$rad \\
  CP-IP distance                 & b=5mm \\
\hline 
Laser parameters & \\
\hline
  Wave length         & $\lambda_L=1.17 \mu{\rm m}$   \\               
  Pulse energy        & 1     J     \\               
  Pulse length        & $\sigma_z^L=0.23 {\rm mm}$     \\
  Peak power density  & 1$\times 10^{18}$ $W/ cm^2$  \\
  Repetition rate     & same as the electron beam \\
  r.m.s spot size     & $\sigma_r^L=2.9\mu{\rm m}$   \\
\hline
\end{tabular}
\end{center}
 \label{tbl:nlc_gg}
\end{table}

With these parameters,  geometric luminosity
of electron-electron collision is 
$8.7\times 10^{33}cm^{-2}s^{-2}$ which is larger
than the typical NLC \epm \ collider
($4.3 \times 10^{33}cm^{-2}s^{-2}$).
Since the luminosity of photon-photon colliders
is approximately proportional to the geometric 
luminosity 
and, unlike \epm \  collider, 
there is no strong beamstrahlung at the interaction, 
the higher geometric luminosity is preferable.  
Laser parameters are chosen so that conversion efficiency 
of incoming electrons in a laser pulse is about 0.65.
The peak laser power density  is about $10^{18} W/ cm^2$ which 
corresponds to nonlinear QED parameter 
$$
\xi ^2=
\left( {{{eE} \over {\omega mc}}} \right)^2
\approx 0.4\left[ {{I \over {10^{18}W/ cm^2}}} \right]
\left[ {{\lambda_L  \over {1.054\mu m}}} \right]^2
\approx 0.4
$$
where $e, E, \omega, m, c, I $ and $\lambda_L$ are
electric charge, strength of laser field, laser photon energy, 
electron mass, speed of light, laser intensity and laser
wave length respectively. 
Here we assumed that the laser profile has a 3-dimensional Gaussian shape
 and it can be focused to diffraction limit.

With this set of electron and laser parameters, the Compton 
kinematic parameter is
$x = 4\E_b\omega/m_e^2=4.47$
and the maximum photon energy $\E_{max}$ in linear Compton limit is
$$
\E_{max} = {x \over x+1}\E_b \approx 200GeV.
$$
which is about $80 \% $ of the original beam energy.

 The treatment of spent electrons coming out from the CP
 is one of the important issues to be considered
 in \gg colliders. 
 If these electrons collide with electrons and photons from 
 the other beam, beam-beam interaction at the IP
 generates low energy electron 
 positron pairs. These pairs are a possible source of 
 detector background as in \epm \ colliders \cite{JLC-I}.
 In this situation, luminosity of \emg and \emm \ collision is 
 comparable to \gg luminosity which could make physics analysis
 complicated.
 For this reason, it is desirable to install magnet 
 between CP and IP to 
 sweep spent electrons away from IP.
 However, the strength of the magnetic field is needed
 to be on the order of 1 Tesla for effective deflection
 of electrons and  
 it is necessary to install the magnet in the very limited space (1cm) 
 between CP and IP.

 In addition, the magnet must not interfere with precise measurement of 
 vertex position of, for example, b quark decay.
 To meet these, 
 much research and development effort is necessary.

 Using NLC simulation parameters, we consider two cases of interaction region
geometry -- without the sweeping magnet between CP and IP, and with it.

Without the magnet, electron beams are collided with $1\sigma_y$  offset
so as to reduce electron beam collision without significantly deteriorating
\gg luminosity. 

The energy spectra of Compton scattered photons are plotted in 
fig.\ref{fig:spec} for linear and nonlinear QED calculations. 
In the simulation, 
it is assumed that the laser beam is 100\% circularly polarized 
and the electron beam is 100\% longitudinally polarized.
The combination of polarization of laser ($P_\gamma$) and  
electron ($P_e$) beams is chosen so that  
$P_\gamma P_e = -1$, which produces a
relatively narrow peak
at  high energy edge.
Comparing nonlinear and linear Compton spectra, 
the maximum energy of photons in nonlinear processes exceeds 
$\E_{max}=x\E_b/(x+1) \approx 200GeV$
due to multiple laser photon absorption.
It is also seen  that the high energy peak of about 200GeV
in linear Compton is shifted to a lower value in nonlinear spectrum.
This is another effect of nonlinear interaction, i.e.,
increasing of effective electron mass.
The peak energy is consistent with the expected value, 
$$
\E_{max}={{x\E_b} \over {x+\xi ^2+1}}\approx 190GeV.
$$

The differential luminosity spectrum
 is shown in fig.\ref{fig:lumgg} for linear
and nonlinear Compton calculations.
In $L_{\gamma \gamma}$ distribution,
high c.m.s energy contriubution is made by collision of Compton 
photons. 
In the low energy region, 
a large low energy tail is seen in the spectrum. 
The source of the tail is  beamstrahlung, i.e.,
collisions of beamstrahlung photons with  
 beamstrahlung and Compton photons.
With nonlinear calculation, high energy peak is
shifted to a lower value due to the shift in Compton
photon spectrum, and the peak becomes 
broader than the linear Compton case.
\gg luminosity in high energy region is about 8\%
of geometric luminosity and 10\% 
in linear Compton calculation because of the 
broadness of the high energy peak.
The nonlinear effect lowers the peak energy 
and broadens the peak; however, with this set of parameters
 $\xi^2=0.4$ and the effect is not very 
significant and is at tolerable level.
Obtained luminosities are summarized in Table 2.

\begin{table}
\begin{center}
\caption{Summary of the luminosity}
\begin{tabular}{|ll|}
\hline
{Linear Compton Simulation} & \\
\hline
  $L_{\gamma \gamma}$   & 0.98$L_{geom}$  \\               
                        & 0.10$L_{geom}$ ($z=W_{\gamma\gamma}/2\E_b>0.65$) \\
  $L_{e \gamma}$        & 0.71$L_{geom}$  \\               
                        & 0.16$L_{geom}$ ($z>0.65$) \\
  $L_{ee}$              & 0.10$L_{geom}$  \\               
                        & 0.05$L_{geom}$ ($z>0.65$) \\
\hline
{Nonlinear Compton Simulation} & \\
\hline
  $L_{\gamma \gamma}$   & 0.88$L_{geom}$  \\               
                        & 0.08$L_{geom}$ ($z>0.65$) \\
  $L_{e \gamma}$        & 0.71$L_{geom}$  \\               
                        & 0.16$L_{geom}$ ($z>0.65$) \\
  $L_{ee}$              & 0.11$L_{geom}$  \\               
                        & 0.06$L_{geom}$ ($z>0.65$) \\
\hline
\end{tabular}
\end{center}
 \label{tbl:lum}
\end{table}

Since there is an  overlap  of electron beams
and of electron 
and photon beams at the interaction point,  
some amount of  $L_{e \gamma}$ 
and $L_{ee}$ is observed. 
From the experimental point of view, the initial 
state of the interaction should be as simple as possible but 
should provide high luminosities at the same time.
These requirements are  conflicting and additional studies
are needed to find an optimum solution.

 The luminosity distribution for the case with sweeping magnet is
 shown in fig.\ref{fig:lumgg_mag}.
 The simulation parameters of the electron and the laser beams are the same
 as in the case without the sweeping magnet 
 except for the distance between CP and IP:
 taking into account comlications of installation of the magnet,
 CP is shifted to 10mm from the IP.
 The strength of magnetic field is 1 Tesla in x direction
 and 250GeV electron is swept 60nm($\approx 6\sigma_y$)
 away in y direction from IP.
 As seen in fig.\ref{fig:lumgg_mag}, \emg and \emm \ 
 luminosities are significantly reduced. 
 Comparing with the non-sweeping magnet case, 
 \gg luminosity is expected to be reduced due to the enlargement of 
 CP-IP distance while it gains a little bit due to
 the absense of $\sigma_y$ offset.
 As a result, \gg luminosity is  
 6\% of geomrtric luminosiry for $z>0.65$ which is slightly smaller than
 non-sweeping case(8\%).

\subsection{Other applications}  

 As the second case study, we applied the program to 
 \epm \ and \emm \ collisions in NLC configuration listed in Table 3 with
 center-of-mass energy $\sqrt{S}=500{\rm GeV}$\cite{NLC}.

\begin{table}
\begin{center}
\caption{Parameters for a \epm and \emm collider} 
\begin{tabular}{|ll|}
\hline
{Electron beam parameters} & \\
\hline
  Beam energy                      & $\E_b$=250{\rm GeV} \\               
  Number of Particles per bunch    & $N=0.65 \times$ $10^{10}$  \\          
  Repetition rate                  & $f_{rep}=180{\rm Hz}$     \\
  Number of bunches per pulse      & $n_b=90$  \\
  Bunch length                     & $\sigma_z$=100$\mu$m \\     
  Bunch sizes (I.P.)               & $\sigma_x$=286nm \\
                                   & $\sigma_y$=4.5nm \\
  Beta functions (I.P.)            & $\beta_x$=8.4mm \\
                                   & $\beta_y$=0.126mm \\               
  Emittances & $\gamma\varepsilon_x$=5.0 $\times$ $10^{-6}$ m$\cdot$rad \\
             & $\gamma\varepsilon_y$=8.0 $\times$ $10^{-8}$ m$\cdot$rad \\
\hline
\end{tabular}
\end{center}
 \label{tbl:nlc_ee}
\end{table}
 The calculated luminosity is shown in fig.\ref{fig:epm}.
 The total \epm \ and \emm \ luminosity is 1.42$L_{geom}$ and 
 $0.55L_{geom}$ respectively.
 As expected, the \epm \ luminosity is enhanced by the collective
 Coulomb interaction ( pinch effect ) while the \emm \ luminosity 
 is reduced to almost half of geometric luminosity due to 
 repulsive coulomb interaction at the IP.

 To simulate \emg collider, the laser pulse should be
 aimed at one  electron beam and the other
 beam should be kept untouched. 
 This simulation is easily set up
 by the combination of \gg and \emm \ parameters and 
 the results are similar to \gg collider without the sweeping magnet.

\section{The Next Step: CAIN 2}
\subsection{Problem in CAIN1.0}
As was demonstrated in the previous section, 
CAIN1.0 can be successfully used for the simulations of general
 linear collider schemes, however 
there are some problems with the structure of the program. 
The main problem comes from the fact that 
IP simulation of CAIN1.0 is essentially the same as ABEL 
which was developed for pure \epm{} simulation.

The IP simulation in CAIN1.0
assumes that each bunch in the initial state consists of 
a single kind of 
particle -- electron or positron with possible mixture of photons. 
(For example, the same distribution is used for particle 
distribution to calculate luminosity and for charge 
distribution to calculate the beam field.)
Although electron positron pairs are created in CP by Breit-Wheeler process, 
the information on the pair particle species is ignored in the IP simulation.
For most of \gg{} collider parameters,
Breit-Wheeler process in CP is kept small and neglecting pair species 
does not affect the simulation significantly.
However, in the case of high laser intensity or high $x$, 
large number of electron positron pairs are created at CP and 
their contribution should not be ignored in the IP simulation. 

It is implicitly assumed in CAIN1.0 that the initial 
energies of electrons/positrons are more or less in the same energy range. 
However, in the case of \gg{} colliders the energy just 
before IP has a wide spread from the full energy down to a few percent. 
This fact makes the various formulas 
(for example the integration of equation of motion) 
adopted in CAIN1.0 somewhat inaccurate. 
In this respect the incoherent pair particles, 
whose energy can be much lower, have no problem 
because they
are treated in a different way. 
However, in fact the spent electrons and 
the incoherent pair particles form an energy spectrum 
almost continuous from a few MeV to hundreds of GeV. 
In this sense there is no reason to treat 
the incoherent pair particles on a different footing.

The orbit angles of incoherent pair particles 
can be as large as hundred milliradians. 
Nevertheless, CAIN1.0 assumes that the $z$-component of 
the velocity is equal to the velocity of light. 
This fact makes the orbit calculation somewhat 
unreliable but it is very hard to modify this point 
in the framework of CAIN1.0.


%

There is another problem which is common for both CP and IP simulations.
As mentioned, the simulation is performed by macro 
particles each of which represents, typically, O($10^6$) real particles. 
If one is interested in the effect of smaller number of particles, 
say, O($10^3$), a very large number of macro particles 
is needed for such a run. 
This drastically affects program speed and required storage. 
There are several ways to avoid this, however. 
One can populate certain regions of the beam 
(the halo, for instance) with macro particles with reduced weight, 
thus enhancing resolution only in the regions of interest. 
Also, the analysis of ``light-weight'' macro particle 
behavior can be done after the collective fields have been calculated, 
thus neglecting the contribution of these particles to the field. 
These methods are not implemented in CAIN1.0. 

\subsection{CAIN2 Project}

In order to overcome the problems stated above, 
the simulation program CAIN2 has been written from scratch 
because the code and memory structure of ABELMOD 
were not adequate for further extension. 
The major differences of the structure of the new version CAIN2 are 
\begin{itemize}
\item All the particles (electron/positron/photon, initial or secondary, etc) 
are stored in the same array and treated on equal basis. 
(Laser beams are not treated as particles: they are `external fields' 
like the field of magnets.)
\item  Instead of invoking separate programs,
various interactions such as laser-Compton, beam-beam interaction
and beam transport are processed one by one at every time step in one
program, if their flags are on. 
This will enable to add new interactions, such as plasma,  
which may take place simultaneously with other interactions. 
\item The new user interface allows much more variety of the configurations 
of the beams and interactions so that applications 
other than linear colliders may be possible. 
For example, one can prepare a neutral beam of mixed \epm{}, 
a bunch consisting of many bunchlets, etc. 
\end{itemize}

The basic assumption in CAIN1.0 is that the collision of the 
two beams is collinear, 
meaning that the crossing angle is very small 
and that each of the two beams, right-going and left-going, 
is a well-defined beam, i.e., 
the mutual angles between the constituents are reasonably small. 
Without this assumption the 
calculation of the beam-beam force would be very complex. 

This requirement has also been adopted in CAIN2 but 
it is relaxed in two respects. 
Firstly, small samples of particles 
(such that their contribution to the beam field is negligible) 
can have large angles. 
This is relevant for incoherent pair particles. 
Secondly, the right-going and left-going beams can make large angles 
so long as each beam is well collimated. 
CAIN2 introduces Lorentz transformation to make the collision collinear. 
Thus, a large crossing angle can be correctly treated. 

The latest version of CAIN2, which is to be completed soon, includes
the following interactions:
\begin{enumerate}
\itemsep 0mm
\item {\label{classical}} Beam deformation by classical field 
(mainly the beam field) 
\item {\label{beamstr}} Quantum-theoretical synchrotron radiation 
(beamstrahlung) 
\item {\label{cohpair}} Coherent pair creation (this was missing in CAIN1.0) 
\item   {\label{linlaser}} Linear interaction of lasers with e$^-$, e$^+$,
$\gamma$.
\item  {\label{nonlinlaser}} Nonlinear interaction of lasers with e$^-$,
e$^+$, $\gamma$.
\item Particle-particle interactions such as the incoherent pair creation 
and bremsstrahlung.
\end{enumerate}
Now, all the processes in CP, TP, and IP can be treated by one program. 
They can be done in a single job or partitioned into separate jobs.

Since the polarization is very important in various applications, 
it is included in most of the above interactions. 
For example, (1) in the above list includes precession in magnetic fields, 
(2), (3), (4) include all possible polarizations, 
and (5) includes longitudinal polarization of all the particles, 
initial and final.

In order to overcome the statistical problem of rare events, 
most interactions now have the `event enhancement factor'. 
For some interactions it is also possible to enhance 
the rate of some part of the spectrum so that, 
for example, one can create more low-energy macro particles (with less weight).

\section{Summary}
 We developed a simulation program 
for phenomena in the interaction regions
of linear colliders which allow us
to estimate realistic luminosity distributions and 
detector backgrounds.
This simulation program can be used for various 
types of linear colliders such as \gg,  \emg, \emm, 
and \epm \ by just switching input parameter.

This program was used for a  
photon-photon collider option of the NLC and 
a realistic luminosity distribution was obtained.
It was also found that nonlinear QED effect is not
negligible in typical parameters for a \gg collider. 
Since particle physics issues as well as  amounts of background events
depend on the luminosity distribution, it gives us 
useful information for further study.

\section*{Acknowledgments}
 We would like to thank Profs.~K.J.~Kim, M.~Ronan and
 Dr.~M.~Xie of LBL for useful discussions.
 Two of the authors (T.T. and T.O.) thank Prof. I.~Endo for 
 his encouragement.

%
%
\noindent
{\Large\bf Figure Captions}
\newcounter{klm}
\begin{list}{Fig.\arabic{klm}}{\usecounter{klm}}
%
%
\item A Schematic view of a photon linear collider.
   \label{fig:schematic}
%
%
\item The scheme of the simulation. The processes and particles
      considered in each step is also shown. 
   \label{fig:structure}
%
%
\item Simulated photon energy spectrum from Compton 
      conversion point without (a) and with(b) nonlinear
      QED effect.
   \label{fig:spec}
%
%
\item Simulated luminosity distribution of a \gg 
    collider without (a) and with (b) nonlinear
    QED effect.
    Solid, dashed and dots line corresponds to \gg, \emg and \emm
 luminosity respectively.
   \label{fig:lumgg}
%
%
\item Simulated luminosity distribution of a \gg 
    collider with sweeping magnet.
    Solid, dashed and dots line corresponds to \gg, \emg and \emm
 luminosity respectively.
   \label{fig:lumgg_mag}
%
%
\item Simulated luminosity distribution of a \epm (a)
   and \emm (b) collider with NLC parameter. 
   \label{fig:epm}

\end{list}

\begin{figure}[h]
\center
\epsfile{file=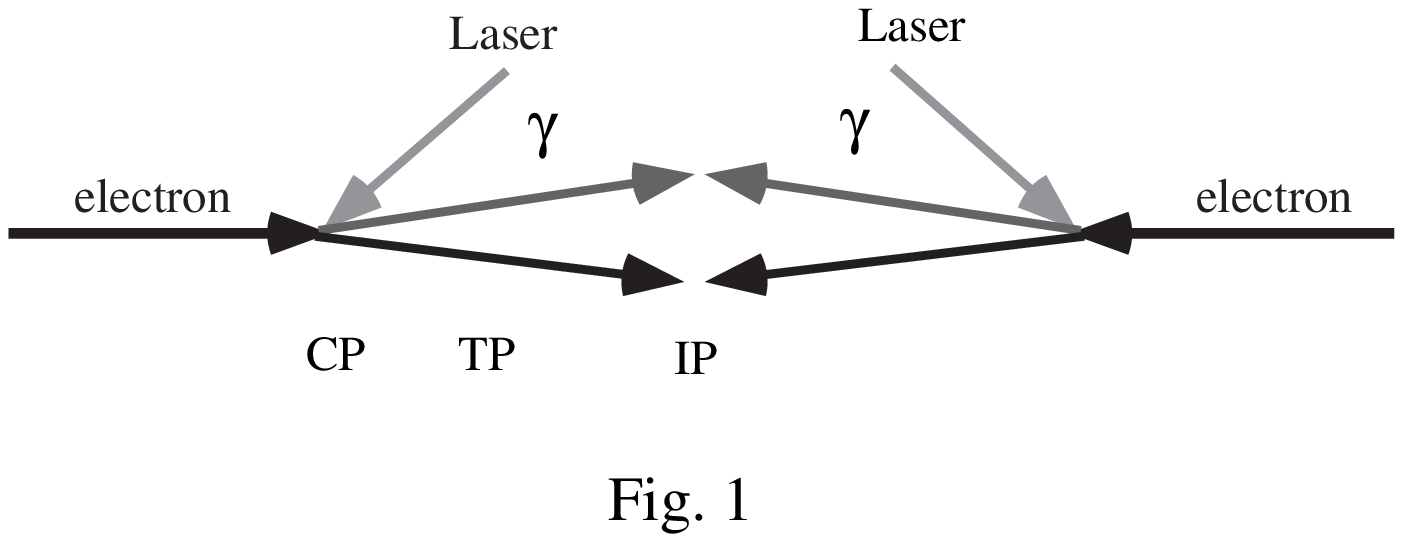,scale=1.0}
\end{figure}

\begin{figure}[h]
\center
\epsfile{file=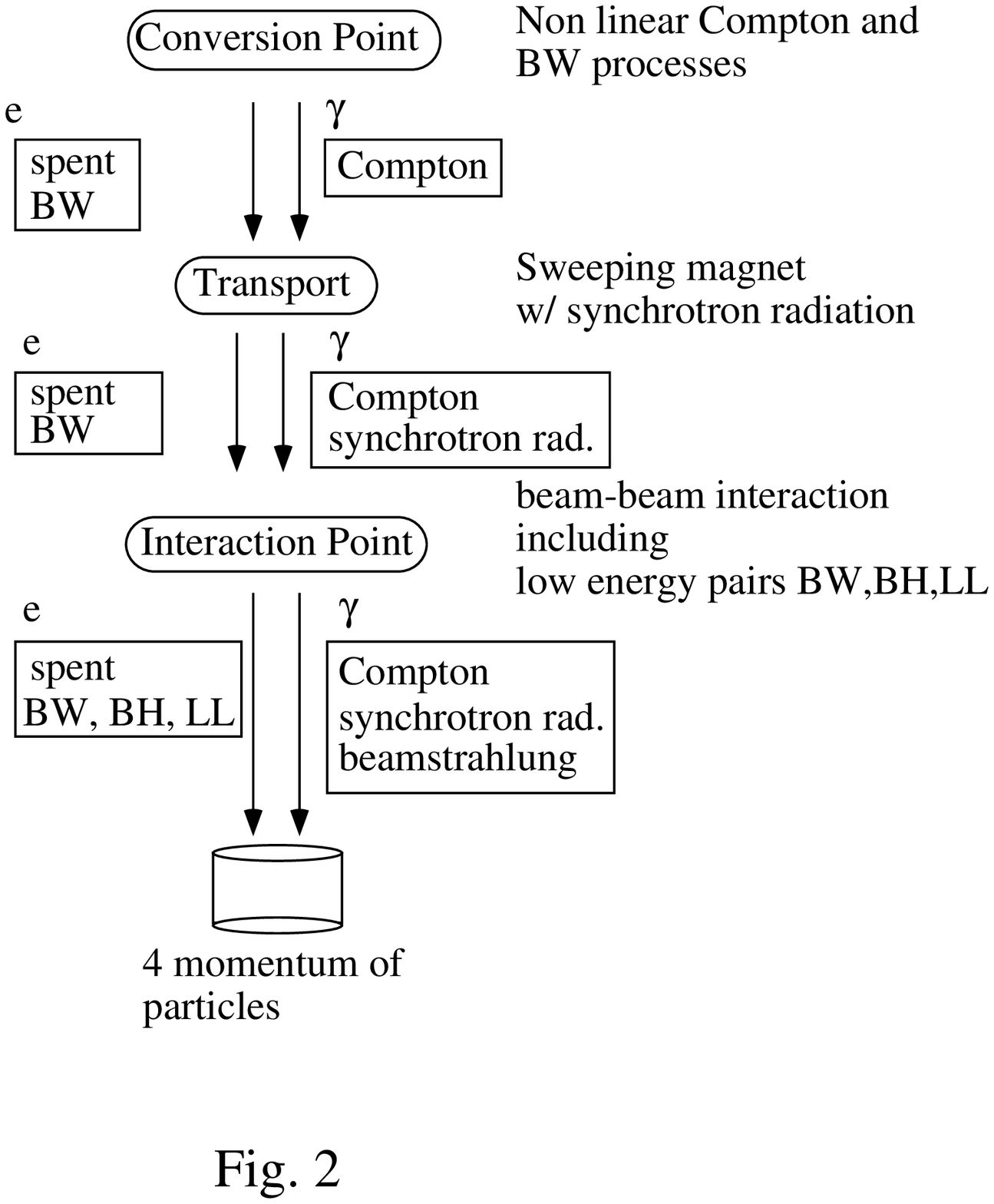,scale=0.9}
\end{figure}

\begin{figure}[h]
\center
\epsfile{file=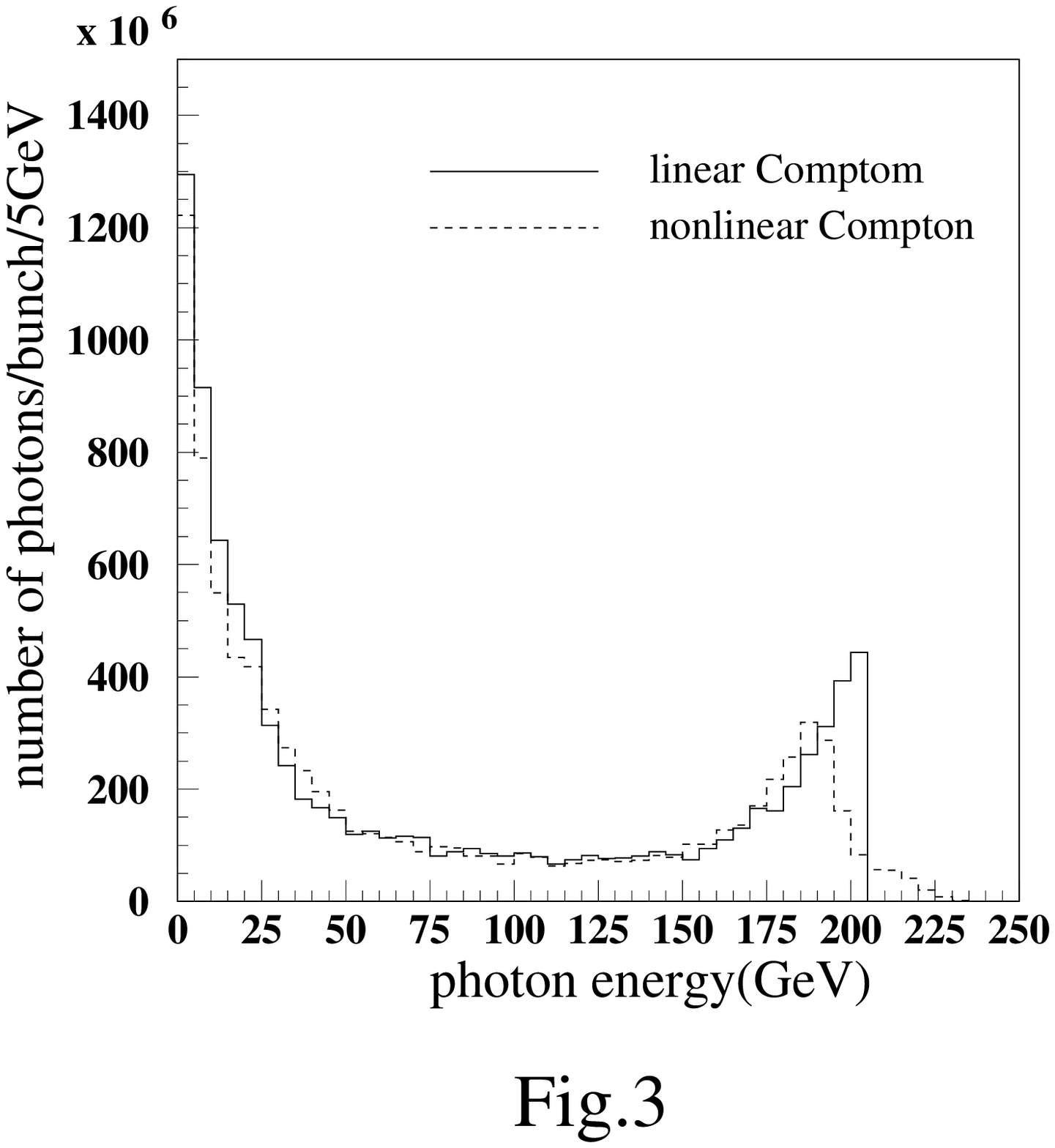,scale=0.75}
\end{figure}

\begin{figure}[h]
\center
\epsfile{file=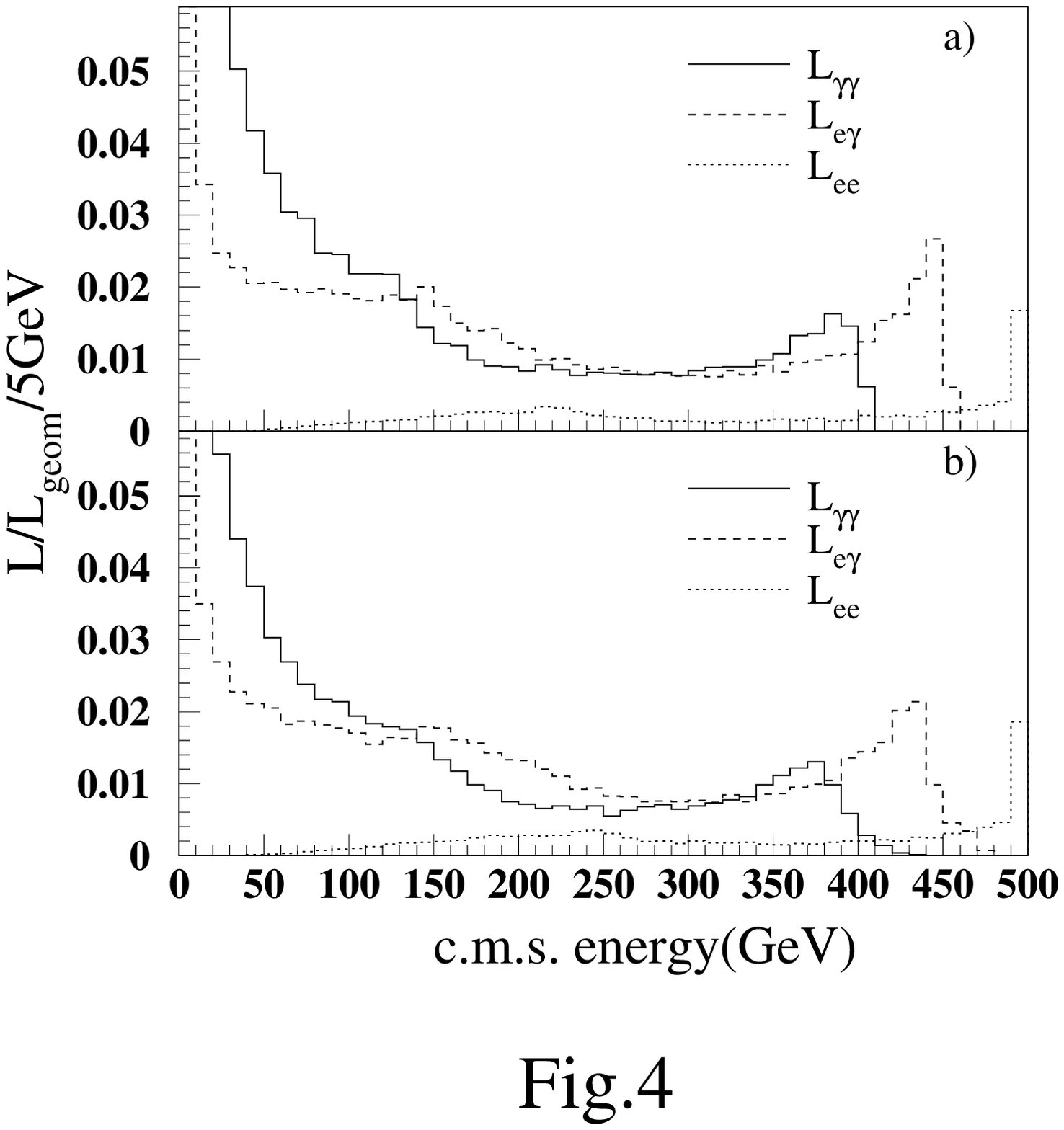,scale=0.75}
\end{figure}

\begin{figure}[h]
\center
\epsfile{file=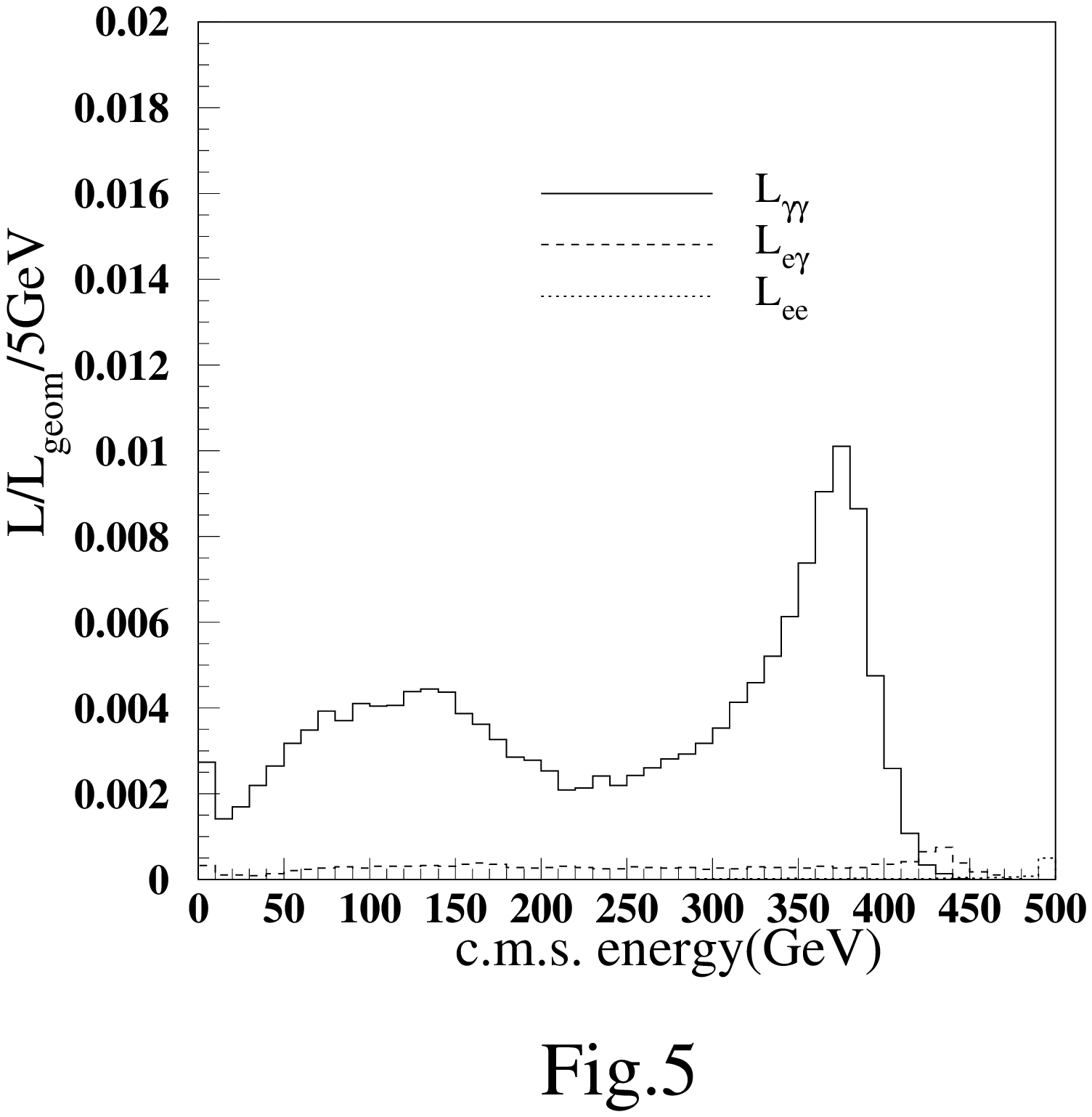,scale=0.75}
\end{figure}

\begin{figure}[h]
\center
\epsfile{file=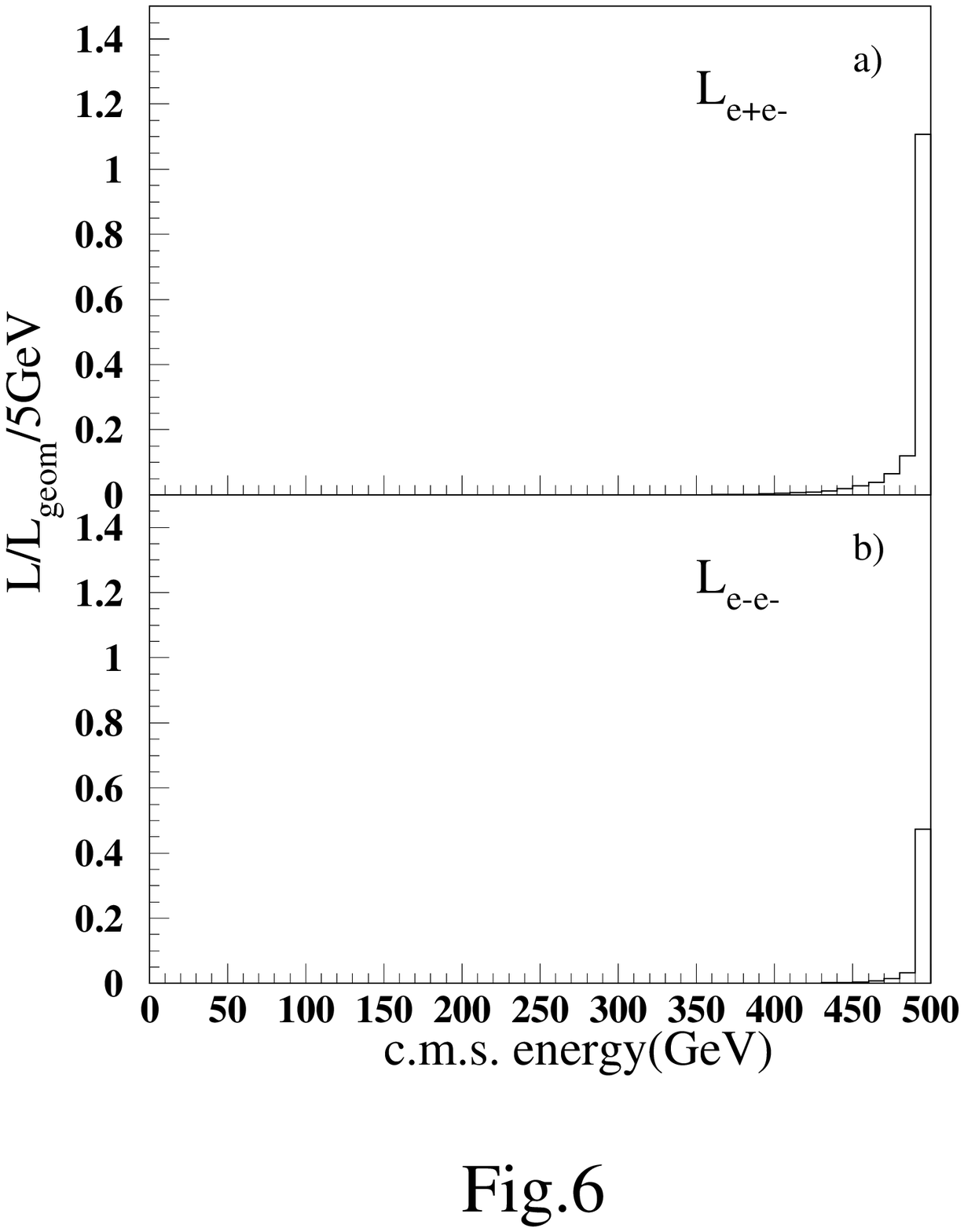,scale=0.75}
\end{figure}


\begin{thebibliography}{99}
\bibitem{ABEL} K.~Yokoya, KEK-Report 85-9 (1985); 
 Nucle. Instr. and Meth. {\bf A251} (1986) 1
\bibitem{pairs} P.~Chen and V.I.~Telnov, Phys. Rev. Lett. {\bf 63} (1989) 1796.
\bibitem{ABELMOD} T.~Tauchi, K.~Yokoya and P.~Chen, Part. Acc. {\bf 41} (1993) 29.
\bibitem{CAIN0} P.~Chen et. al., Nucl. Instr. Meth.{\bf A335} (1995) 107.
\bibitem{Ohgaki} T.~Ohgaki and T.~Takahashi,
 Nucle. Instr. Meth. {\bf A373} (1996) 185 
\bibitem{volkov} D.M.~Volkov, Z. Phys. {\bf 94} (1935) 250
\bibitem{tsai} Y.S.~Tsai Phys. Rev. D{\bf 48}(1993) 96
\bibitem{JLC-I} JLC-I KEK-Report 92-16 (1992)
\bibitem{NLC} Zeroth-Order Design Report for the Next Linear Collider,
		appendix B. SLAC-474
\end{thebibliography}
\end{document}